\documentclass[12pt,twoside]{article}
\usepackage[mathscr]{eucal}
\usepackage{amsmath,amsfonts,amssymb,amsthm}
\usepackage[matrix,arrow]{xy}
\usepackage{mathabx}
\usepackage{amsthm,amsmath,latexsym,amssymb,amsfonts,amsbsy}
\usepackage{wrapfig}
\usepackage{graphicx}
\usepackage{float}

\usepackage[breaklinks]{hyperref}
\usepackage{amsmath}
\usepackage{amssymb}
\usepackage{braket}
\usepackage[T1]{fontenc}
\usepackage{sidecap}

\usepackage{tikz,pgf}
\usetikzlibrary{shapes}
\usetikzlibrary{calc}
\usetikzlibrary{decorations.pathmorphing}
\usetikzlibrary{decorations.pathreplacing,shapes.misc}
\usetikzlibrary{positioning}
\usetikzlibrary{arrows}
\usetikzlibrary{decorations.markings}

\usetikzlibrary{intersections}

\def\be{\begin{equation}}
\def\ee{\end{equation}}
\def\ba{\begin{array}}
\def\ea{\end{array}}

\def\dps{\displaystyle}

\def\A{a}
\def\B{b}

\def\a{\tilde{1}}
\def\b{\tilde{2}}
\def\1{\tilde{1}}
\def\2{\tilde{2}}
\def\3{\tilde{3}}

\newdimen\tableauside\tableauside=1.0ex
\newdimen\tableaurule\tableaurule=0.4pt
\newdimen\tableaustep
\def\phantomhrule#1{\hbox{\vbox to0pt{\hrule height\tableaurule
width#1\vss}}}
\def\phantomvrule#1{\vbox{\hbox to0pt{\vrule width\tableaurule
height#1\hss}}}
\def\sqr{\vbox{%
  \phantomhrule\tableaustep

\hbox{\phantomvrule\tableaustep\kern\tableaustep\phantomvrule\tableaustep}%
  \hbox{\vbox{\phantomhrule\tableauside}\kern-\tableaurule}}}
\def\squares#1{\hbox{\count0=#1\noindent\loop\sqr
  \advance\count0 by-1 \ifnum\count0>0\repeat}}
\def\tableau#1{\vcenter{\offinterlineskip
  \tableaustep=\tableauside\advance\tableaustep by-\tableaurule
  \kern\normallineskip\hbox
    {\kern\normallineskip\vbox
      {\gettableau#1 0 }%
     \kern\normallineskip\kern\tableaurule}%
  \kern\normallineskip\kern\tableaurule}}
\def\gettableau#1 {\ifnum#1=0\let\next=\null\else
  \squares{#1}\let\next=\gettableau\fi\next}

\renewcommand{\tilde}{\widetilde}

\newcommand{\bref}[1]{\textbf{\ref{#1}}}

\newcommand{\np}{I_{+-}^{(2)}}
\newcommand{\nm}{I_{-+}^{(2)}}
\newcommand{\lp}{I_{++}^{(2)}}
\newcommand{\lm}{I_{--}^{(2)}}
\newcommand{\kp}{I_{+-}^{(3)}}
\newcommand{\km}{I_{-+}^{(3)}}
\newcommand{\tp}{I_{++}^{(3)}}
\newcommand{\tm}{I_{--}^{(3)}}

\def\cF{\mathcal{F}}

\def\cL{\mathcal{L}}

\def\cO{\mathcal{O}}

\numberwithin{equation}{section} \makeatletter
\@addtoreset{equation}{section}

\def\be{\begin{equation}}
\def\ee{\end{equation}}
\def\ba{\begin{array}}
\def\ea{\end{array}}
\def\d{\partial}
\def\dps{\displaystyle}

\def\ba{\begin{array}}
\def\ea{\end{array}}
\def\d{\partial}
\def\dps{\displaystyle}

\begin{document}

\vspace{5mm}
\begin{flushright}
FIAN-TD-2015-08 \\
\end{flushright}

\vspace{10mm}

\begin{center}
{\Large\textbf{Monodromic vs geodesic  computation  \\ 
\vspace{5mm}
of Virasoro classical  conformal  blocks}}

\vspace{10mm}

{\large Konstantin  Alkalaev$^{\;a,b}$ and   Vladimir Belavin$^{\;a,c}$}

\vspace{7mm}

\textit{$^{a}$I.E. Tamm Department of Theoretical Physics, \\P.N. Lebedev Physical
Institute,\\ Leninsky ave. 53, 119991 Moscow, Russia}

\vspace{0.3cm}

\textit{$^{b}$Moscow Institute of Physics and Technology, \\
Dolgoprudnyi, 141700 Moscow region, Russia}

\vspace{0.3cm}

\textit{$^{c}$Department of Quantum Physics, \\ 
Institute for Information Transmission Problems, \\
 Bolshoy Karetny per. 19, 127994 Moscow, Russia}

\vspace{0.5cm}

\thispagestyle{empty}


\end{center}
\begin{abstract}
We compute 5-point classical conformal blocks with two heavy, two light, and one superlight operator using the monodromy approach up to  third order in the superlight expansion. By virtue of the AdS/CFT correspondence we show the equivalence of the resulting expressions to those obtained in the bulk  computation for the corresponding geodesic configuration.

\end{abstract}
%
%

\section{Introduction}

Conformal blocks represent important elements of any conformal field theory \cite{Belavin:1984vu}. In general, they
are certain functions of conformal dimensions $\{\Delta_i\}$ and Virasoro central charge $c$ which are  completely defined by the conformal symmetry. In the limit where $c$ goes to infinity they are approximated by the so-called  classical conformal blocks. 
Recently, a remarkable interpretation of the classical conformal blocks in the context of the AdS$_3$/CFT$_2$ correspondence has	 been investigated \cite{Fitzpatrick:2014vua,Asplund:2014coa,Caputa:2014eta,Fitzpatrick:2015zha,Hijano:2015rla,Alkalaev:2015wia,Hijano:2015qja}. 
It was shown that some class of classical CFT$_2$ conformal block can be described by means of a particular classical mechanics in AdS$_3$.  

There exist different types of the classical conformal blocks depending on the behaviour of the conformal dimensions as $c\rightarrow \infty$  \cite{Zamolodchikov:1995aa}. One can distinguish between two limiting cases: conformal blocks with only heavy and only light operators. In this note we focus on the case of two heavy operators of equal  conformal dimensions producing  in the bulk either a conical defect or  BTZ black hole  \cite{Fitzpatrick:2014vua,Asplund:2014coa,Fitzpatrick:2015zha,Hijano:2015rla}.
The  bulk geodesic configuration corresponding to the  $n$-point block with only two heavy fields
consists of $2n-5$ massive scalar particles propagating in the  background geometry produced by the heavy operators \cite{Alkalaev:2015wia}. 
It was pointed out in \cite{Hijano:2015zsa,Hijano:2015qja} that
these configurations are nothing but the ordinary Witten exchange diagrams with the difference that there is no integration over positions of the vertices of the geodesic graph. 

While the $4$-point conformal block in the heavy-light approximation\footnote{For the general 4-point block the basic computation tool is Zamolodchikov recursion relations \cite{Zamolodchikov:1985ie, Zamolodchikov:1987ie}, for the recent development see \cite{Perlmutter:2015iya}. Alternative recursion in the important case of the vacuum conformal blocks   was proposed recently in \cite{Fitzpatrick:2015foa}. For multi-point blocks the AGT combinatorial representation is applicable (at finite value of the central charge in rational and non-rational CFTs with Virasoro symmetry see \cite{Alday:2009aq, Bershtein:2014qma, Alkalaev:2014sma}.}  was explicitly computed \cite{Fitzpatrick:2015zha,Hijano:2015rla} an exact consideration of the $n$-point case is hampered by technical difficulties related to solving associated higher order algebraic equations. To analyze  multi-particle configurations we proposed to use an additional approximation procedure   with respect to a small parameter which can be chosen to be a mass of one of the particles whose worldline  ends on the boundary  (\textit{i.e.}, it is a conformal dimension of one of  external fields) \cite{Alkalaev:2015wia}. This allows to iteratively reconstruct $n$-point heavy-light classical conformal block starting from the $(n-1)$-point one. Such a {\it super-light} approximation  applies when there is a known expression for the $n$-point heavy-light conformal block. This is the case with the $5$-point heavy-light block  considered as a deformation of the exactly known heavy-light $4$-point  block  with respect to a small classical conformal dimension of the third light operator. In this paper we use this  procedure to analyse the bulk/boundary correspondence of the conformal block/geodesic  Witten diagram  computation of the $5$-point heavy-light block in the sub-leading orders of the super-light expansion.

Our computation of the $5$-point  classical conformal block on the boundary relies on the study of the monodromy properties of the auxiliary Fuchsian differential equation. It is reduced to the computation of the so-called  accessory parameters which are partial derivatives of the conformal block function. In the bulk we use the geodesic approach involving a different but related  set of quantities  -- angular momenta of external and intermediate geodesic segments, and the mechanical  action of the geodesic configuration or, equivalently, the geodesic length.  Analogously to the accessory parameters, the external angular momenta are defined as derivatives of the mechanical action  which in turn is related to the classical conformal block \cite{Hijano:2015rla}.  We expect that the geodesic description can be successively derived from the monodromy approach by finding counterparts of the monodromy approach constituents on the bulk side. In this paper we compare the systems of algebraic equations describing both the accessory and angular parameters and find out that they are generally different but have a common  physically relevant root which leads to the conformal block function.

The paper is organized as follows. In Section  \bref{sec:monodrom} we apply the monodromy approach
to find equations on the accessory parameters for the classical $5$-point conformal block. In Section \bref{sec:algeq} we formulate the super-light approximation procedure and  compute the accessory parameters up to the third order in the dimension of the super-light operator. In Section \bref{sec:bulk}  we perform  the corresponding bulk computation. In Section \bref{sec:adscft} we discuss  explicit relations between the classical conformal block and the geodesic length, and between the accessory parameters and external angular momenta. Then we discuss a relation between two systems of  algebraic equations on accessory/angular parameters.  Using the super-light expansion method we find a perturbative solution to the bulk equations on angular parameters up to the third order. 
In Section \bref{sec:block} we compute the 5-point classical conformal block and the geodesic length up to the third order in the superlight dimension and find out  exact match of two answers. Section \bref{sec:concl} contains our conclusions.

\section{Classical conformal block and monodromy problem}
\label{sec:monodrom}

An $n$-point conformal block function $\cF(z_i, \Delta_i, \tilde \Delta_j)$  is a holomorphic contribution to the correlation function of $n$ primary fields in points $z_i$ coming from a given set of  Virasoro representations in the intermediate channels. In addition to the central charge $c$ and  conformal weights $\Delta_i$ of the  external operators  the conformal block depends on  conformal weights $\tilde \Delta_j$ in the intermediate channels. For the spherical topology the fusion channel is  represented by the corresponding  fishbone diagram. 

\begin{figure}[H]
\centering
\begin{tikzpicture}


\draw [line width=1pt] (30,0) -- (32,0);
\draw [line width=1pt] (32,0) -- (32,2);
\draw [smooth, tension=1.0, line width=1pt, decorate, decoration = {snake, segment length = 2mm, amplitude=0.4mm}] (32,0) -- (34,0);
\draw [line width=1pt] (34,0) -- (34,2);
\draw [smooth, tension=1.0, line width=1pt, decorate, decoration = {snake, segment length = 2mm, amplitude=0.4mm}] (34,0) -- (36,0);
\draw [line width=3pt] (36,0) -- (36,2);
\draw [line width=3pt](36,0) -- (38,0);


\draw (29.3,-0) node {$0, \Delta_1$};
\draw (32,2.5) node {$z_2, \Delta_{2}$};
\draw (34,2.5) node {$z_{3}, \Delta_{3}$};
\draw (36,2.5) node {$1, \Delta_{h}$};
\draw (38.8,0) node {$\infty, \Delta_h$};


\fill (30,0) circle (0.8mm);

\fill (32,0) circle (0.8mm);

\fill (34,0) circle (0.8mm);
\fill (32,2) circle (0.8mm);

\fill (36,0) circle (0.8mm);
\fill (34,2) circle (0.8mm);

\end{tikzpicture}
\caption{The $5$-point classical heavy-light conformal block. Two bold lines on the right represent heavy operators.
As usual the projective invariance  is used to fix  three insertion positions as $z_1 =0$,  $z_4= 1$, $z_5 = \infty$.}
\label{5block}
\end{figure}
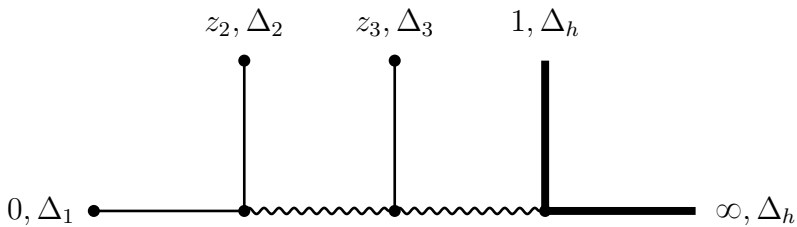

In the classical limit  $c\rightarrow \infty$ the conformal blocks  are exponentiated (see, \textit{e.g.}, \cite{Zamolodchikov1986,Harlow:2011ny}) as
\be
\label{ccb}
\cF(z_i, \Delta_i, \tilde \Delta_j) = \exp\big[-\frac{c}{6}f(z_i, \epsilon_i, \tilde \epsilon_j)\big]\;,
\ee
where $\epsilon_i = \Delta_i/c$ and $\tilde \epsilon_j = \tilde \Delta_j/c$ are called respectively external and intermediate classical conformal dimensions 
and $f(z_i, \epsilon_i, \tilde \epsilon_j)$ is the main object of our study --  classical conformal block. To compute  the classical conformal block in the $5$-point case  we apply the monodromy method  (see, \textit{e.g.}, \cite{Zograf,Takhtajan:1994vt} for a general discussion). From now on, a function $f(z_i)$ denotes  the $5$-point classical  block
related to the quantum block  with the diagram presented on the  Fig. \bref{5block} (we omit conformal dimensions as they always remain the same).

The  $5$-point conformal block can be obtained by considering an auxiliary $6$-point correlation function  
$\langle V_{12}(z) V_1(z_1) \cdots V_5(z_5)\rangle$, 
where 
$V_i(z_i)$ are primary operators with dimensions $\Delta_i$, $i=1,...,5$ and $V_{12}(z)$ is the  second level
degenerate operator. Below we use the standard  Liouville   parametrization   $c=1+6(b+b^{-1})^2$  \cite{Zamolodchikov:1995aa}.
Because of the  decoupling condition $(b^{-2}L_{-1}^2+L_{-2}) V_{12}(z)=0$  the $6$-point correlation function obeys  the following
second order differential equation  \cite{Belavin:1984vu}
\be
\label{decouple}
\Big[\frac{1}{b^2} \frac{\d^2}{\d z^2} + \sum_{i=1}^5 \Big(\frac{\Delta_i}{(z-z_i)^2}+ \frac{1}{z-z_i}\frac{\d}{ \d z_i}\Big)\Big] \langle V_{12}(z) V_1(z_1) \cdots V_5(z_5)\rangle = 0\;.
\ee 
In the classical limit $b \rightarrow 0$ the $6$-point auxiliary correlation function 
behaves as
\be
\label{ccb5}
\langle V_{12}(z) V_1(z_1) \cdots V_5(z_5)\rangle \,\Big |_{b \rightarrow 0} \rightarrow \psi(z) \exp(b^{-2} f(z_i))\;,
\ee
where the exponential factor $f(z_i)$ is the sought 5-point classical conformal block and
the auxiliary holomorphic function $\psi(z)$ is governed by a Fuchsian equation  arising from the above decoupling condition 
\be
\label{master}
\frac{d^2 \psi(z)}{dz^2}  + T(z) \psi(z) = 0\;.
\ee
Here $T(z)$  is the holomorphic component of the classical stress-energy tensor 
\be
\label{Tz}
T(z) = \sum_{i=1}^5 \Big(\frac{\epsilon_i}{(z-z_i)^2} +  \frac{c_i}{z-z_i}\Big)\;,
\ee
where $\epsilon_i$ are the classical conformal dimensions and $c_{i}$ are the accessory parameters related to the classical conformal block 
as follows
\be
\label{cbgrad}
c_i (z) = \frac{\d f(z)}{\d z_i}\;,\qquad i = 1,...,5\;.
\ee
The asymptotic behaviour $T(z) \sim z^{-4}$ at infinity implies that the accessory parameters satisfy the following linear conditions 
\be
\label{moeq}
\sum^5_{i = 1} c_i = 0\;,
\qquad
\sum^5_{i = 1} (c_i z_i + \epsilon_i) = 0\;, 
\qquad
\sum^5_{i = 1} (c_i z_i^2 + 2\epsilon_i z_i) = 0\;.
\ee
Fixing $z_1 = 0, z_{4} = 1, z_5 = \infty$ the above conditions are solved as 
\be
\label{ccc}
\ba{c}
c_1 = -c_2(1-z_2) - c_3(1-z_3) + \epsilon_1 + \epsilon_2 + \epsilon_3 + \epsilon_4 - \epsilon_5\;,
\\
\\
c_4 = - c_2 z_2 - c_3 z_3-\epsilon_1 - \epsilon_2 - \epsilon_3 - \epsilon_4 + \epsilon_5 \;,
\qquad
c_5 = 0\;, 
\ea
\ee
where accessory parameters $c_2$ and $c_3$ are taken as independent variables. Using \eqref{ccc}   function $T(z)$ takes the form 
\be
\label{Tred}
\ba{l}
\dps
T(z) = \frac{\epsilon_1}{z^2}   +\frac{\epsilon_2}{(z-z_2)^2} + \frac{\epsilon_3}{(z-z_3)^2}+\frac{\epsilon_4}{(z-1)^2}+
 
\\
\\ 
\dps
\hspace{4cm}+\, c_2\frac{z_2 (z_2 - 1)}{z(z - z_2) (z - 1)}  + c_3 \frac{ z_3 (z_3 - 1)}{z(z - z_3) (z - 1)} - \frac{\epsilon_1 + \epsilon_2 + \epsilon_3 +\epsilon_4 -\epsilon_5 }{z (z - 1)}\;.
\ea
\ee
 
\subsection{Heavy-light approximation} 
 
Classical conformal blocks in the heavy-light limit contain two heavy operators while others are light \cite{Fitzpatrick:2014vua}. Let  $\epsilon_{4} = \epsilon_{5}\equiv \epsilon_h$ be the dimension of two heavy fields, while fields  with dimensions $\epsilon_1, \epsilon_2, \epsilon_{3}$ be light. It means that the dimension of heavy operators is fixed in the semiclassical limit while those of light operators tend to zero. The  equation \eqref{master} can then be solved perturbatively. Let us expand all functions as
\be
\label{expan}
\ba{l}
\psi(z) = \psi^{(0)}(z)+ \psi^{(1)}(z) +\psi^{(2)}(z)+ ... \;,
\\
\\
T(z) = T^{(0)}(z)+ T^{(1)}(z)+T^{(2)}(z)+ ... \;,
\\
\\
c_2(z) = c_2^{(0)}(z)+ c_2^{(1)}(z) +c_2^{(2)}(z)+ ... \;,
\\
\\
c_3(z) = c_3^{(0)}(z)+ c_3^{(1)}(z) +c_3^{(2)}(z)+ ... \;,

\ea
\ee
where  expansion parameters are light conformal dimensions. We assume that the accessory parameter expansion starts with terms linear in the conformal dimensions so that $c_{2,3}^{(0)}= 0$. 
In the case of the heavy-light conformal blocks it is sufficient to consider just the first order corrections (see \cite{Fitzpatrick:2014vua,Hijano:2015rla} for the discussion of the 4-pt case). The  lowest order equations are then 
\be
\label{deceq}
\ba{l}
\dps
\Big(\frac{d^2}{dz^2}   + T^{(0)}(z)\Big) \psi^{(0)}(z) = 0\;,
\qquad \quad
\Big(\frac{d^2}{dz^2}   + T^{(0)}(z)\Big) \psi^{(1)}(z) =  - T^{(1)} \psi^{(0)}(z) \;,
\ea
\ee
where the stress-energy tensor components are directly  read off from \eqref{Tred},
\be
\label{Tpert}
\ba{l}
\dps
T^{(0)}(z) = \frac{\epsilon_h}{(z-1)^2}\;,
\qquad\qquad

 T^{(1)}(z) = \frac{\epsilon_1}{z^2}   +\frac{\epsilon_2}{(z-z_2)^2} + \frac{\epsilon_3}{(z-z_3)^2}+
 
\\
\\ 
\dps
\hspace{5cm}+\, c_2\frac{z_2 (z_2 - 1)}{z(z - z_2) (z - 1)}  + c_3 \frac{ z_3 (z_3 - 1)}{z(z - z_3) (z - 1)} - \frac{\epsilon_1 + \epsilon_2 + \epsilon_3 }{z (z - 1)}\;.
\ea
\ee
Here and below we use notation $c_2(z)$ for $c_2^{(1)}(z)$ and $c_3(z)$ for $c_3^{(1)}(z)$ as they are the only coefficients which will come in further calculations.

The two branches in the zeroth order are given by 
\be
\label{zos}
\psi_{\pm}^{(0)}(z) = (1-z)^{\gamma_{_\pm}}\;,
\qquad
\gamma_{\pm} = \frac{1\pm\alpha}{2}\;, 
\qquad
\alpha  = \sqrt{1-4 \epsilon_h}\;.
\ee
Using the method of variation of parameters and noting that the Wronskian of the zeroth order solutions is $W(z) = \alpha$ we find  the first order corrections     
\be
\label{fos}
\ba{c}
\dps
\psi^{(1)}_{+}(z) =  \frac{1}{\alpha}\psi_{+}^{(0)}(z) \int dz\, \psi^{(0)}_-(z)T^{(1)}(z)\psi_{+}^{(0)}(z)   - \frac{1}{\alpha} \psi_{-}^{(0)}(z) \int dz\, \psi^{(0)}_+(z)T^{(1)}(z)\psi_{+}^{(0)}(z)\;,  
\\
\\
\dps
\psi^{(1)}_{-}(z) = \frac{1}{\alpha}\psi_{+}^{(0)}(z) \int dz\, \psi^{(0)}_-(z)T^{(1)}(z)\psi_{-}^{(0)}(z)   - \frac{1}{\alpha} \psi_{-}^{(0)}(z) \int dz\, \psi^{(0)}_+(z)T^{(1)}(z)\psi_{-}^{(0)}(z)\;.

\ea
\ee
Since both $\psi_{\pm}^{(0)}(z)$ and  $T^{(1)}(z)$ have poles then $\psi_{\pm}^{(1)}(z)$ has  branch points identified with the punctures at $z_2$ and $z_3$. 

\subsection{Contour integration and  monodromy}

To find the monodromy we evaluate the following  integrals
\be
\ba{c}
\dps
\qquad I_{++}^{(k)} = \frac{1}{\alpha}\oint_{\gamma_k} dz\, \psi^{(0)}_-(z)T^{(1)}(z)\psi_{+}^{(0)}(z)\;,
\qquad
I_{+-}^{(k)} = -\frac{1}{\alpha}\oint_{\gamma_k} dz\, \psi^{(0)}_+(z)T^{(1)}(z)\psi_{+}^{(0)}(z)\;,
\\
\\
\dps
I_{-+}^{(k)} = \frac{1}{\alpha}\oint_{\gamma_k} dz\, \psi^{(0)}_-(z)T^{(1)}(z)\psi_{-}^{(0)}(z)\;,
\qquad
I_{--}^{(k)} = -\frac{1}{\alpha}\oint_{\gamma_k} dz\, \psi^{(0)}_+(z)T^{(1)}(z)\psi_{-}^{(0)}(z)\;,
\ea
\ee
over two contours $\gamma_{2}$ and $\gamma_3$ enclosing points $\{0, z_2\}$ and $\{0,z_2,z_3\}$ respectively. 
We find 
\be
\ba{l}
\dps
\np  = + \frac{2\pi i }{\alpha}\big[+\alpha \epsilon_1 +c_2(1-z_2) - \epsilon_2 +c_3(1-z_3) - \epsilon_3-(1-z_2)^\alpha[c_2(1-z_2) - \epsilon_2(1+\alpha)] \big]\;,
\\
\\
\dps
\nm  = -\frac{2\pi i }{\alpha}\big[-\alpha \epsilon_1  +c_2(1-z_2) - \epsilon_2 +c_3(1-z_3) - \epsilon_3-(1-z_2)^{-\alpha}[c_2(1-z_2) - \epsilon_2(1-\alpha)] \big]\;,
\\
\\
\dps
\lp  = -\lm = \frac{2\pi i }{\alpha}\big[c_3(1-z_3)- \epsilon_3\big]\;,
\\
\ea
\ee
and 
\be
\ba{l}
\dps
\kp  = +\frac{2\pi i }{\alpha}\big[+\alpha \epsilon_1 + c_2(1-z_2) - \epsilon_2 +c_3(1-z_3) - \epsilon_3-
\\
\\
\hspace{23mm}-(1-z_2)^{\alpha}[c_2(1-z_2) - \epsilon_2(1+\alpha)] -(1-z_3)^{\alpha}[c_3(1-z_3) - \epsilon_3(1+\alpha)]  \big]

\;,
\\
\\
\dps
\km  = -\frac{2\pi i }{\alpha}\big[-\alpha \epsilon_1 + c_2(1-z_2) - \epsilon_2 +c_3(1-z_3) - \epsilon_3-
\\
\\
\hspace{23mm}-(1-z_2)^{-\alpha}[c_2(1-z_2) - \epsilon_2(1-\alpha)] -(1-z_3)^{-\alpha}[c_3(1-z_3) - \epsilon_3(1-\alpha)]  \big]\;,
\\
\\
\dps
\tp  = \tm = 0\;.
\\
\\
\ea
\ee

Within the heavy-light expansion two monodromy matrices $\mathbb{M}=\{M_{ij},\;i,j=\pm\} $ associated with the contours $\gamma_2$ and $\gamma_3$  
\be
\begin{pmatrix}
   \psi_{+}(z) \\
   \psi_{-}(z) 
\end{pmatrix}
\rightarrow 
\begin{pmatrix}
   M_{++} & M_{+-} \\
   M_{-+} & M_{--} 
\end{pmatrix}
\begin{pmatrix}
   \psi_{+}(z) \\
   \psi_{-}(z) 
\end{pmatrix}
\ee
are expanded as 
\be
\mathbb{M} = \mathbb{M}_0 + \mathbb{M}_1 + \mathbb{M}_2 + ...\;.  
\ee
The first order $\mathbb{M}_0$ defines the monodromy of $\psi^{(0)}(z)$. 
As the  contours do not enclose the point $z=1$ it follows that $\mathbb{M}_0 = \mathbb{I}$.  
In the linear order the monodromy matrices are given by  
\be
\mathbb{M}({\gamma_2}) = \begin{pmatrix}
   1+ \lp & \np \\
   \nm & 1-\lp 
\end{pmatrix}\;,
\qquad\quad
\mathbb{M}(\gamma_3) = \begin{pmatrix}
   1 & \kp \\
   \km & 1 
\end{pmatrix}\;.
\ee 

On the other hand, the monodromy matrices over contours $\gamma_2$ and $\gamma_3$ are defined by the conformal dimensions  of the fields in the intermediate channels. Up to similarity transformation they are 
\be
\mathbb{\tilde M}(\gamma_2) =  - \begin{pmatrix}
   e^{+\pi i \Lambda_1} & 0 \\
   0 & e^{-\pi i \Lambda_1} 
\end{pmatrix}\;,
\qquad
\mathbb{\tilde M}(\gamma_3) =  - \begin{pmatrix}
   e^{+\pi i \Lambda_2} & 0 \\
   0 & e^{-\pi i \Lambda_2} 
\end{pmatrix}\;,
\ee
where 
$\Lambda_1 = \sqrt{1-4\tilde \epsilon_1}$ and 
$\Lambda_2 = \sqrt{1-4\tilde \epsilon_2}$ parametrize  intermediate dimensions. Computing corresponding eigenvalues  we arrive at the following equation system
\be
\label{m1}
\sqrt{\lp\lp + \np\nm}  = 2\pi i \,\tilde \epsilon_1\;,
\quad \qquad
\sqrt{\kp \km}  = 2 \pi i \, \tilde \epsilon_2\;.
\ee

\section{Solving the monodromic equations}
\label{sec:algeq}

In what follows we solve the monodromic equations \eqref{m1} for the special choice of the parameters
\be
\label{econs}
\epsilon_1 = \epsilon_2\;,
\qquad
\tilde\epsilon_1 = \tilde\epsilon_2\;.
\ee 
Equations \eqref{m1}  being squared take  the form 
\be
\label{32}
\ba{l}
\dps
\big(X+(1-a)Y\big) X= a I^2 \;,
\\
\\
\dps
\big(X+(1-b)Y +  b E\big)\big(b X+ a(1-b) Y + a E\big)= a b I^2 \;,
\ea
\ee
where
\be
\label{ch6}
X = (1-a)(x - \epsilon_1) +(1+a)\alpha\epsilon_1\;,
\qquad
Y = (y - \epsilon_3)\;,
\qquad
E =\alpha\epsilon_3\;,
\qquad
I = \alpha \tilde \epsilon_1\;,
\ee
and 
\be
\label{origch}
\ba{c}
x = (1-z_2)c_2\;,
\quad
y = (1-z_3)c_3\;,
\\
\\
\A = (1-z_2)^\alpha\;,
\qquad
\B = (1-z_3)^\alpha\;.
\ea
\ee
Thus, the monodromic equation system   consists of two quadratic equations on two accessory parameters. Following the B\'{e}zout's theorem  the system  has at most four different solutions (more precisely, none, two or four). Each equation describes a conic and their intersection points (at most four) are the solutions. The first equation in \eqref{32} is easily solved as $Y = Y(X)$ and thereby isolating variable $X$ one arrives at a fourth order equation that can be solved explicitly using  the well-known formulas for roots. However, the corresponding expressions containing square and cubic radicals are too complicated. One can  also check that the discriminant is not zero so that all roots are different.

Exact solutions to the monodromic system are yet to be found. Instead, we propose to use an expansion which treats a solution as a deformation of some seed solution with respect to one of  the conformal dimensions. As a seed solution we take the 4-point accessory parameter while the deformation parameter is the third conformal dimension $\epsilon_3$. Thus, the $5$-point accessory parameters are expanded as  
\be
\label{perser}
\ba{l}
c_2 = c_2^{(0)} + \epsilon_3 c_2^{(1)}+\epsilon^2_3 c_2^{(2)}+ \cdots  \;,
\\
\\
c_3 = \epsilon_3 c_3^{(1)}+\epsilon^2_3 c_3^{(2)}+ \cdots  \;,
\ea
\ee   
where the zeroth-order $c_2^{(0)}$ is the $4$-point accessory parameter \cite{Fitzpatrick:2014vua,Hijano:2015rla}, while $c^{(k)}_{2,3}$ are corrections, $k=1,2,...\;$ (see our notational  comment below \eqref{Tpert}). Taking $\epsilon_3 = 0$ yields  $c_3 = 0$ so that  two equations \eqref{32} are reduced to a single equation
\be
\label{quadr}
X^2  - aI^2 = 0\;.
\ee 
Recalling variable changes  \eqref{origch} and \eqref{ch6}  we solve the above equation as  
\be
\label{fitz_sol}
X = \pm \sqrt{a} \alpha \tilde \epsilon_1\;: \qquad x  = \epsilon_1 + \frac{(a+1)}{(a-1)}\alpha \epsilon_1 \pm \alpha \frac{\sqrt{a}}{(a-1)} \tilde  \epsilon_1\;,
\qquad c^{(0)}_2 = \frac{x}{1-z_2}\;,
\ee
where the conformal block asymptotics fixes the sign to be $+$.


Having exact solution $c_2^{(0)}$ and using the expansion procedure one can easily generate higher order corrections to find series \eqref{perser}. The point is that substituting the power series into the monodromic equations we find out  that the zeroth order term satisfies the  quadratic equation \eqref{quadr} while all higher order terms are subjected to some linear equations. In terms of variables $x$ and $y$ \eqref{origch} the perturbation series is 
\be
\label{pertubxyH}
x = \sum_{n=0}^\infty \epsilon^n_3 x_n\;,
\qquad \quad
y = \sum_{n=1}^\infty \epsilon^n_3 y_n \;,
\ee
where $x_0$ is the exact seed solution \eqref{fitz_sol}. All corrections up to the third order are given by  
\be
\hspace{-5cm}
\ba{l}
\dps
x_0=\epsilon_1+\epsilon_1\alpha\frac{(a+1)}{(a-1)}+\tilde{\epsilon}_1\alpha \frac{\sqrt{a}}{a-1}\;,
\qquad \quad x_1=\frac{\alpha}{2}\,\frac{a+b^2}{a-b^2}\;,
\\
\\
\dps
x_2= \frac{\alpha}{2\tilde{\epsilon}_1} \bigg[\frac{b\sqrt{a}(a-2 a b+b^2)(a-2 b+b^2)}{(a-b^2)^3}+\frac{(a-1)(a+b^2)^2}{4\sqrt{a}(a-b^2)^2}\bigg]\;,
\\
\\
\dps
x_3= \frac{\alpha}{2\tilde{\epsilon}_1^2}\bigg[\frac{a b(b-1)(a-2 ab +b^2)(a-2 b+b^2)(a-3 a b+3b^2-b^3)}{(a-b^2)^5}\bigg]\;,
\label{xlist}
\ea
\ee
and
\be
\ba{l}
\dps
y_1=1-\alpha\frac{a + b^2}{a - b^2}\;,
\qquad\quad
y_2=\frac{\alpha}{\tilde{\epsilon}_1}\bigg[\frac{b\sqrt{a} (-a + 2 a b - b^2) (a - 2 b + b^2) }{(a -b^2)^3}\bigg] \;,
\\
\\
\dps
y_3=\frac{\alpha}{2\tilde{\epsilon}_1^2}\bigg[\frac{b (a - 2 a b + b^2) (a - 2 b + b^2) (a^2 + a^3 - 8 a^2 b + 
    6 a b^2 + 6 a^2 b^2 - 8 a b^3 + b^4 + a b^4)}{(a - b^2)^5 }\bigg]\;.
\label{ylist}
\ea
\ee
Corrections $c^{(k)}_{2,3}$ at  $k=1,2,...$ in \eqref{perser} can be reproduced using  the variable change \eqref{origch}. 

\section{Bulk geodesic approach }
\label{sec:bulk}

In this section we shortly discuss the basic ingredients of the multi-particle mechanics in the bulk  used to compute classical conformal blocks in the $4$-point case \cite{Hijano:2015rla}  and in the $n$-point case  \cite{Alkalaev:2015wia}. Let us consider a massive point particle propagating in the three-dimensional space with a conical singularity. The corresponding metric is given by       
\be
\label{metric}
ds^2  = \frac{\alpha^2}{\cos^2 \rho}\Big( dt^2 +\sin^2\rho d\phi^2 +\frac{1}{\alpha^2} d\rho^2 \Big)\;,
\ee
where  $t\in \mathbb{R}$, $\rho \in \mathbb{R}^+$, $\phi \in [0,2\pi)$. The angle deficit is  $2\pi(1-\alpha)$ where $\alpha \in [0,1]$. Here, $\alpha = 1$ corresponds to the global $AdS_3$ space,  $\alpha = 0$ corresponds to the BTZ black hole threshold, while $\alpha = 1/n,$ where $n \in \mathbb{Z}_+$ gives $AdS_3/\mathbb{Z}_n$ conifold.    
The angle deficit metric \eqref{metric} describes the constant negative curvature space with the Ricci scalar $R =-8$, while  the conformal boundary  is reached at $\rho \rightarrow \pi/2$. 

The action associated with the interval  \eqref{metric} reads  
\be
\label{OPA}
S = \epsilon\int_{\lambda^{'}}^{\lambda^{''}} d\lambda \, \sqrt{g_{tt} \dot{t}^2+g_{\phi\phi} \dot{\phi}^2+g_{\rho\rho} \dot{\rho}^2}\;, 
\ee 
where $\epsilon$ is a classical conformal dimension identified with a mass, the metric coefficients are read off from \eqref{metric}, $\lambda$ is the  evolution parameter and the dot stands for the corresponding  derivative. As the action \eqref{OPA} is reparameterization invariant the evolution parameter can be conveniently chosen so that the proper velocity is unit. We observe  that the boundary coordinates $\phi$ and $t$ are cyclic, that is the corresponding Lagrange function has no explicit time and angular dependence
\be
\label{cyclic}
\frac{\d \cL}{\d \phi} = \frac{\d \cL}{\d t} = 0\;,
\ee 
and therefore both  angular and time momenta are conserved $\dot p_\phi = \dot p_t = 0$. The particle dynamics can be reduced to the level surface characterized  by  constant values of the conserved momenta. However, the corresponding Rauth function is rather complicated. It can be simplified by taking either $p_\phi =0$ and/or $p_t=0$ in which case the Rauth function is just the Lagrange function \eqref{OPA} at $\dot \phi = 0$ and/or $\dot t = 0$.  

In our case we are interested in $2n-5$ massive particles subjected to particular boundary conditions meaning that $n-2$ of their worldlines puncture the conformal boundary at fixed points $w_1,...,w_{n-2}$, where $w = \phi+i t$. In general, each particle travels on geodesic characterized by its own angular and time momenta. There are two natural configurations of boundary attachment points when either time or angular momenta of all particles can be simultaneously set to zero. In the $t = const$ case the surface level is a space-like disk with a number of attachment points placed on the boundary circle. Heavy operators can be visualized as sitting  in the center of the disk.  In the $\phi = const$ case the surface level is a time-like infinite strip with a number of attachment points placed on the boundary line. Heavy operators are on the opposite edge at infinitely separated points. In each case both external and intermediate worldlines have different constant angular/time momenta according to their attachment points and vertex positions.   

We note that both the disk and the infinite strip turn out to be constant negative curvature two-dimensional surfaces  with equal Ricci scalars $R = -2$. Both metrics can be cast into the Poincar\'{e} disk form $d \tilde s^2  = \frac{1}{4} \frac{d u d \bar u}{(1-u\bar u)^2}$, where $u, \bar u$ are images of the complexified  coordinates $\phi\pm i\rho$ or $\rho \pm i t$, and the conformal boundary is mapped to the circle  $u\bar u=1$. It follows that the dynamics in both slices can be described quite uniformly. In particular, the length of a given geodesic graph in a constant angle slice can be explicitly computed and be shown that it is related to the length of the same graph in a constant time slice by virtue of  Wick rotation of  attachment points coordinates.

In the sequel we consider a fixed time slice with the boundary attachments arranged circle-wise. In this case the Rauth function describes a massive particle moving in the disk 
$d\tilde s^2  = \frac{d\rho^2 + \alpha^2 \sin^2 \rho\, d \phi^2}{\cos^2 \rho}$.
The geodesic length of a given segment characterized by a constant angular parameter 
$s \equiv |p_\phi|/\alpha$ is the disk line element integrated along a given path 
\be
\label{lambda}
S = \epsilon  \ln \frac{ \sqrt{\eta}}{\sqrt{1+\eta} +  \sqrt{1 - s^2 \eta}}\,\Bigg|_{\eta^{'}}^{\eta^{''}}\;,
\ee
where $\eta^{'} = \cot^2 \rho^{'}$ and $\eta^{''} = \cot^2 \rho^{''}$ are initial/final radial positions. As the metric blows up at the conformal boundary $\rho = \pi/2$ the length is to be  regularized by introducing a near boundary cutoff.  Then, the geodesic length of a multi-line graph is obtained by summing \eqref{lambda} with various  angular parameters.  On the other hand, the angular parameters are uniquely fixed by conditions that follow from extremizing the total action with fixed boundary attachments.  

As an example we consider a geodesic between two boundary points $0$ and $\phi$. Using \eqref{lambda} one finds that up to the boundary cutoff the geodesic length is given by
$S(\phi) = \ln\sin (\alpha \phi/2)$.
As the angular coordinate $\phi$ runs from $0$ to $2\pi$ we find out that the geodesic is the arc stretched  between the boundary points. \footnote{Let us note that function $S(\phi)$ is periodic with the period $4\pi/\alpha$ which is greater than the period of the angular variable $\phi$. This gives rise to the interesting case of the so-called "long geodesics" which are the curves stretched between the two points and winding around the conical singularity. In the case of the conifold $AdS_3/\mathbb{Z}_n$ they are the images of the true geodesic lines in $AdS_3$ and compute the  entwinement   \cite{Balasubramanian:2014sra}. It would be crucial to consider the long geodesics  as new elements of $n$-point bulk graphs and find an interpretation  in terms of the boundary conformal theory.} The arc along with a radial geodesic line and bulk-to-bulk geodesic segments  are the basic elements of any $n$-point geodesic graph. 

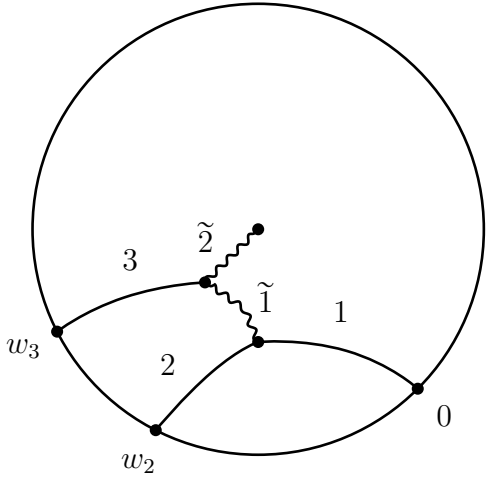
\begin{SCfigure}
\begin{tikzpicture}[line width=1pt]
\draw (0,0) circle (3cm);

\foreach \a in {1,2,...,40}{
\draw (\a*360/40: 3.5cm) coordinate(N\a){};
\draw (\a*360/40: 3cm) coordinate(K\a){};
\draw (\a*360/40: 2cm) coordinate(L\a){};
\draw (\a*360/40: 1.5cm) coordinate(I\a){};
\draw (\a*360/40: 1.1cm) coordinate(J\a){};
\draw (\a*360/40: 1cm) coordinate(M\a){};
;}


\draw plot [smooth, tension=1.0, line width=1pt] coordinates {(K35) (L34) (I30)};
\draw plot [smooth, tension=1.0, line width=1pt] coordinates {(K27) (L28) (I30)};
\draw plot [smooth, tension=1.0, line width=1pt] coordinates {(K23) (L23) (M25)};


\draw [smooth, tension=1.0, line width=1pt, decorate, decoration = {snake, segment length = 2mm, amplitude=0.4mm}] (M25)  -- (0,0);


\draw  [smooth, tension=1.0, line width=1pt, decorate, decoration = {snake, segment length = 2mm, amplitude=0.4mm}]  (I30) -- (J29) -- (M25);

\fill (I30) circle (0.8mm);
\fill (M25) circle (0.8mm);

\fill (K35) circle (0.8mm);
\fill (K27) circle (0.8mm);
\fill (K23) circle (0.8mm);
\fill (0,0) circle (0.8mm);

\draw (1.1,-1.1) node {$1$};
\draw (-1.2,-1.8) node {$2$};
\draw (-1.7,-0.4) node {$3$};
\draw (-0.7,-0.1) node {$\b$};
\draw (0.1,-0.9) node {$\a$};
\draw (N23) node {$w_3$};
\draw (N27) node {$w_2$};
\draw (N35) node {$0$};

\end{tikzpicture}
\caption{A constant time  disk. Solid lines $1,2,3$ represent external operators with dimensions $\epsilon_i$, $i=1,2,3$,  wavy lines $\a,\b$ represent intermediate operators with dimensions $\tilde \epsilon_j$, $j=1,2$. } 
\label{5bulk}
\end{SCfigure}

The geodesic graph in the $5$-point case consists of five segments  three of which have boundary attachments $0, w_2, w_3$ and meet two inner segments in two vertices, see Fig. \bref{5bulk}. Positions of the  vertices are defined through the equilibrium conditions on the  angular and radial momenta 
\be
\label{rels}
\tilde s_2 = 0\;,
\qquad
\epsilon_3 s_3 - \tilde \epsilon_1 \tilde s_1 = 0\;,
\qquad
\epsilon_1 s_1 - \epsilon_2 s_2 - \tilde \epsilon_1 \tilde s_1 = 0\;,
\ee
\be
\label{vertcoord11}
\epsilon_1\sqrt{1-s_1^2 \eta_1} + \epsilon_2\sqrt{1-s_2^2 \eta_1} = \tilde\epsilon_1\sqrt{1-\tilde s_1^2  \eta_1}\;,
\qquad
\epsilon_3\sqrt{1-s_3^2 \eta_2} + \tilde \epsilon_1\sqrt{1-\tilde s_1^2 \eta_2} = \tilde \epsilon_2\;,
\ee
and the angular conditions saying that angular separations between the first and second lines and between the first and the third lines are $w_2$ and $w_3$, respectively,  
\begin{eqnarray}
\label{firsteq}
e^{i\alpha w_2}=
\frac{\big(\sqrt{1-s_1^2\, \eta_1}-i s_1 \,\sqrt{1+\eta_1}\big)\big(\sqrt{1-s_2^2\, \eta_1}-i s_2\, \sqrt{1+\eta_1}\big)}
{(1-i s_1) (1-i s_2) }\;,
\end{eqnarray}
\vspace{1mm}
\be
\label{secondeq}
e^{i\alpha w_3}= \frac{\big(\sqrt{1-s_3^2 \eta_2 }-i s_3 \sqrt{1+\eta_2}\big)
\big(\sqrt{1-\tilde s_1^2 \eta_2}-i \tilde s_1 \sqrt{1+\eta_2}\big)\big(\sqrt{1-s_1^2 \eta_1}-i s_1\sqrt{1+\eta_1}\big)}{(1-i s_3)\big(\sqrt{1-\tilde s_1^2 \eta_1}-i \tilde s_1\sqrt{1+\eta_1}\big)(1-i s_1)}\;,
\ee
where $s_1, s_2, s_3 $ and $\tilde s_{1}, \tilde s_{2}$ are respectively external and intermediate angular parameters, and $\eta_{1}$ and $\eta_2$ are radial positions of the vertices. In total, we have seven algebraic (irrational) equations on seven variables.

Modulo regulator terms the total multi-particle  action is given by a sum of geodesic segments  
\be
\label{actionFULL1}
S(w_2, w_3) =  \epsilon_1 L_1(w_2, w_3) + \epsilon_2 L_2(w_2, w_3)+ \epsilon_3 L_3(w_2, w_3) +\tilde\epsilon_1 L_{\a}(w_2, w_3)+ \tilde\epsilon_2 L_{\b}(w_2, w_3) \;,
\ee
where    
\be
\label{LLLLL}
\ba{c}
\dps
L_{1,2} =  - \ln \frac{\sqrt{\eta_1}}{\sqrt{1+\eta_1}+\sqrt{1  - s_{1,2}^2\eta_1}}\;,
\qquad
L_{3} = - \ln \frac{\sqrt{\eta_2}}{\sqrt{1+\eta_2}+\sqrt{1- s_3^2\eta_2}}\;,
\\
\\
\dps
L_{\a} = \ln \frac{\sqrt{\eta_1}}{\sqrt{1+\eta_1}+\sqrt{1- \tilde s_1^2\eta_1}}- \ln \frac{\sqrt{\eta_2}}{\sqrt{1+\eta_2}+\sqrt{1- \tilde s_1^2\eta_2}} \;,
\quad
L_{\b} = \ln \frac{\sqrt{\eta_2}}{1+ \sqrt{1+\eta_2}}\;.
\ea
\ee
Solving equations \eqref{rels} - \eqref{secondeq} we find all variables as functions of boundary attachment positions $w_1, ..., w_{n-2}$ and all classical conformal dimensions. 
However, exact solutions are not known yet and therefore we shall use the expansion method which computes a solution as a perturbative series over some exactly known seed solution. 

\section{Monodromy vs geodesic approach}

\label{sec:adscft}

Finding the geodesic length can be viewed from a slightly different perspective. As the geodesic length is just the multi-particle action with particular boundary conditions given by fixed attachment points of external lines one finds out that corresponding canonical momenta are partial derivatives of the action with respect to boundary attachment coordinates. The resulting system of partial differential equations can be explicitly integrated so that its solution certainly coincides with geometric computation of the geodesic length given in Section \bref{sec:bulk}.

It follows that both the monodromy and the geodesic methods deal with potential vector fields whose components are identified with accessory or external angular parameters, while potentials are identified with respectively the classical conformal block or the geodesic length. Given  vector field components as functions of puncture coordinates we easily find the corresponding potential function. However, finding the components which are subjected to algebraic equation system turns out to be quite complicated problem. Within the monodromy approach such a system arises from studying  monodromy properties of the conformal block while within the geodesic approach analogous system follow from studying geometric characteristic of geodesic segments. Below we analyze the respective algebraic equation systems on the both sides and show that they have a common root while the systems themselves are different. The two systems are then related by virtue of a combined transformation that can be viewed as the AdS/CFT correspondence.

Let us consider the  action \eqref{OPA} describing a geodesic segment attached somewhere to the boundary. The on-shell variation is $\delta S  = p_\mu \delta x^{\mu}$, where $p_\mu$ is the momentum in the attachment point, while $\delta x^{\mu}$ is the  variation of the boundary coordinate. In our case there are three boundary attachments $w_1 = 0$ and $w_2,w_3$ (see Fig. \bref{5bulk}) so that 
\be
\label{Swvar}
\alpha \epsilon_2 s_2(w_2, w_3) = \frac{\d S(w_2, w_3)}{\d w_2}\;,
\qquad
\alpha \epsilon_3 s_3(w_2, w_3) = \frac{\d S(w_2, w_3)}{\d w_3}\;.
\ee
As noted  in \cite{Hijano:2015rla} in  the $4$-point case the angular momenta relations \eqref{Swvar} are analogous  to the definition of the accessory parameters \eqref{cbgrad}. Indeed, 
the accessory parameters are defined in much the same way as 
\be
\label{cbgrad2}
\ba{l}
\dps
c_2 (z_2,z_3) = \frac{\d f(z_2,z_3)}{\d z_2}\;,
\qquad
c_3 (z_2,z_3)  = \frac{\d f(z_2,z_3)}{\d z_3}\;,

\ea
\ee
cf. \eqref{cbgrad}.
Note that there are  the following consistency conditions \footnote{It is worth noting that the first condition can be represented as the Cauchy-Euler type equation 
$a\frac{\partial y}{\partial a} =b \frac{\partial x}{\partial b}$,  
where $a,b$ and $x,y$ are given by \eqref{origch}. Recalling  expressions for the accessory parameters \eqref{xlist} and \eqref{ylist} we see that this is indeed the case in each particular order of the $\epsilon_3$ expansion.}
\be
\label{conscon}
\frac{\d c_3}{\d z_2} - \frac{\d c_2}{\d z_3}=0\;,
\qquad
\epsilon_3\frac{\d s_3}{\d w_2} - \epsilon_2\frac{\d s_2}{\d w_3}=0\;.
\ee 
Formally, the two systems above define potential vector fields in two dimensions which  can be related to each other by means of a coordinate transformation along with a particular transformation of the respective potentials. A transition from the boundary insertion points  $z_{1,2,3}$ to the  attachment points $w_{1,2,3}$ in the bulk is given by  
\be
\label{coord}
\ba{c}
\dps
w_m  = i \ln (1-z_m)\;,\qquad m = 1,2,3\;.
\ea 
\ee
In particular, $z_1 = 0$ maps to $w_1 = 0$, cf. Fig. \bref{5block} and \bref{5bulk}.
Naturally, \eqref{coord} maps the complex plane to the cylinder interpreted as conformal boundary of the asymptotically AdS$_3$ space. The transformation  is diagonal in the sense of the corresponding diagonal Jacobian matrix.  The multi-particle action and the classical conformal block are related as
\be
\label{fS}
f(z_2,z_3) = S(w_2,w_3)+ i\epsilon_2 w_2 + i\epsilon_3 w_3\;, 
\ee
where coordinates are related by \eqref{coord}.
We find that three accessory  parameters and three external angular parameters are related as
\be
\label{cs}
c_m = \epsilon_m \, \frac{1\pm i \alpha s_m(w)}{1-z_m}\;,\qquad m = 1,2,3\;,
\ee
where $\pm$ depends on the direction of the $s_m$ flow. According to Fig. \bref{5bulk} and relations \eqref{rels} momenta $s_2,s_3$ and  $s_1$ have opposite signs in \eqref{cs}, it is $-$ and $+$ respectively. Note that the accessory parameter $c_1$ is implicit within the monodromic approach since  only two of the accessory parameters are left independent, cf. \eqref{ccc}. On the other hand, relations \eqref{rels} say that $\epsilon_1 s_1  = \epsilon_2 s_2 + \epsilon_3 s_3$, and using \eqref{cs} it is exactly mapped  to \eqref{ccc}.



As discussed above both accessory parameters and angular momenta  are  subjected to  algebraic equation systems \eqref{moeq}, \eqref{m1} and \eqref{rels}-\eqref{secondeq}. Within the monodromy approach there are five variables $c_1, ..., c_5$ (accessory parameters) subjected to  three linear and two irrational equations $M_\alpha(c) = 0$, $\; \alpha = 1,...,5$. Within the geodesic approach there are seven  variables $s_1, s_2, s_3, \tilde s_1, \tilde s_2$ (external/intermediate angular momenta) and $\eta_1 , \eta_2$ (radial vertex positions) subjected to three linear and four  irrational  equations
$G_{I}(s,\tilde s, \eta) = 0$, $\;I = 1,...,7$. 
In principle, one might expect that after isolating two independent bulk variables (say, $s_2$ and $s_3$) the residual two geodesic equations should match exactly with  the monodromic equations \eqref{m1} provided \eqref{cs}. However, this  is not the case. Instead, a weaker version of the equivalence turns out to be true -- the systems are required to have at least one common root. 

For future reference, recalling the definition of quantities $a$ and $b$ \eqref{origch} and introducing angle positions  $\theta_2 = \alpha w_2/2$ and $\theta_3 = \alpha w_3/2$ we find out that two types of variables are related  as
\be
\label{foot1}
a =\exp(-2 i \theta_2)\;, \qquad b= \exp(-2 i \theta_3)\;.
\ee
Parameters $\theta_{2,3}$ proved to be convenient when describing solutions of the bulk equations and now we see that they are directly related to the quantities $a$ and $b$  naturally introduced within the monodromic framework.


\subsection{4-point case}

The geodesic equations read off from the general system in the 4-pt case with  coinciding conformal dimensions  $\epsilon_1 = \epsilon_2$ \eqref{econs} are given by  
\be
\label{originalEQB4}
2\epsilon_1  \sqrt{1-s^2\, \eta } = \tilde\epsilon_1\;,
\qquad
(\sqrt{1-s^2\, \eta }-i s \sqrt{1+\eta})= \frac{(1-i s)}{\sqrt{a}}\;,
\ee
where we denoted $s_1 = s_2 \equiv s$. The first equation here expresses $\eta$ as a function of $s$. Isolating independent variable $s$ one can show that the second equation has the polynomial representation 
\be
\label{quadraticEQB}
(s+i)(s+ i \frac{(a+1) - \sqrt{a} \varkappa}{1-a}) = 0\;, \qquad \quad\varkappa = \frac{\tilde \epsilon_1}{\epsilon_1}\;.
\ee 
We see that modulo pre-factor $i \alpha \epsilon_1$ the second root of this residual geodesic equation coincides with the root \eqref{fitz_sol} of the monodromic quadratic equation \eqref{quadr}, \textit{i.e.}, $x -\epsilon_1 = - i \alpha \epsilon_1 s$, where $x$ is related to the  accessory parameter $c_2$ via  redefinition \eqref{origch}. The final relation between $c_2$ and $s$ is exactly \eqref{cs}.

It is instructive to have both monodormic \eqref{quadr} and geodesic \eqref{quadraticEQB} equations expressed in the same notation. By making use of the change \eqref{cs} the bulk and boundary quadratic equations are uniformly given by 
\be
\label{vbbv}
\ba{l}
\dps
\text{Monodromic equation: }\qquad \Big(s+ i \frac{(a+1) - \sqrt{a} \varkappa}{1-a}\Big)^2 = 0\;,
\\
\\
\dps
\;\;\text{Geodesic  equation: }\qquad\;(s+i)(s+ i \frac{(a+1) - \sqrt{a} \varkappa}{1-a}) = 0\;.

\ea
\ee
Indeed, we  see that the above equations do not coincide  but have a common root. Note  the second root in the geodesic equation $s = -i$. Recalling that $s$ defines the angular momentum as $p_\phi = \alpha s$ we find out that the second root yields a pure imaginary momentum $p_\phi = -i \alpha$.     
The corresponding vertex does not admit an original geometric interpretation since the radial position $\rho$ defined by variable $\eta = \cot^2 \rho$ from \eqref{originalEQB4} is also imaginary by virtue of the constraints  for classical conformal dimensions \cite{Alkalaev:2015wia}. Nonetheless, it would be interesting to study both formal and  physical meaning of the corresponding solution for $S(w)$ and $f(z)$  elsewhere.


\subsection{5-point case }

A polynomial representation of the geodesic equations in the 5-point case analogous to \eqref{quadraticEQB} is unknown yet. Instead, from the previous discussion we learn that computing all the roots is superfluous. Analyzing the first  angular equation \eqref{firsteq} we obtain exact  results, while the second angular equation  \eqref{secondeq} can be considered only perturbatively by virtue of  the expansion method used to find solutions to the monodromic equations. In particular, we show that, roughly speaking, the first and the second  monodromic equations in \eqref{32} follow from respectively the first and the second  angular equations.     
Indeed, seven bulk variables include two vertex positions which can be explicitly solved in terms of five external and intermediate angular parameters, $\eta_{1,2} = \eta_{1,2}(s, \tilde s|\epsilon, \tilde \epsilon)$. Then, the remaining five geodesic equations include three linear relations \eqref{rels} and two irrational angular equations \eqref{firsteq} and \eqref{secondeq}. This structure is analogous to that of the monodromic equations which include three linear equations \eqref{moeq} and two irrational equations  \eqref{m1} on five accessory parameters. 

\vspace{-4mm}

\paragraph{Exact relations.} By a direct computation we show that the second equation in \eqref{vertcoord11} and the first angular equation \eqref{firsteq} can be conveniently represented as a third order  equation on variables $s_1+s_2$ and $s_1 s_2$ \cite{Alkalaev:2015wia}.  Two of three roots should be discarded while the remaining root yields the following constraint 
\be
\label{512}
s_2  = \frac{\sigma +  \cos 2 \theta_2 +  s_1 \sin 2 \theta_2}{-s_1 + s_1 \cos 2 \theta_2 - \sin 2 \theta_2}\;,\qquad\quad \sigma = 1  - \frac{\tilde \epsilon_1^2}{2\epsilon_1^2}\;,
\ee
while relations \eqref{rels}  in this case reduce to  
$s_1 = s_2+\frac{\epsilon_3}{\epsilon_1} s_3$.
Equation \eqref{512} is  the second order equation in  variables $s_2$ and $s_3$. It is not equivalent to the first angular equation in the bulk but instead represents one of its roots. 
Parameterizing coefficients in \eqref{512} as 
$\cos 2\theta_2  = (1+a^2)/2a$, 
$\sin 2\theta_2  = (1-a^2)/2ia$, and $\cot \theta_2 = i (1+a)/(1-a)$ (see \eqref{foot1}),  
and denoting 
\be
\label{change2}
\ba{l}
X = - i\alpha \epsilon_1(1-a) (s_2 + \cot \theta_2)\;,
\qquad
Y = -i \alpha \epsilon_3 s_3\;,
\ea
\ee
we show that quadratic equation 
\eqref{512} can be  cast into the form 
\be
\label{321}
X^2 +(1-a) XY -aI^2 = 0\;,
\ee
which is exactly the first monodromic equation \eqref{32}, while the coordinate changes \eqref{change2} and \eqref{origch} yield relation \eqref{cs}. In this way we see that the equivalence can be achieved just for roots of the algebraic equations and not for the  equations themselves.

\vspace{-4mm}

\paragraph{Expanding over the root.} The remaining part of the geodesic equations is given by the first vertex equation in \eqref{vertcoord11} and the second angular equation \eqref{secondeq} which is much more complicated than the first angular equation. For convenience, we use the first angular equation \eqref{firsteq} to represent the second angular equation \eqref{secondeq} as follows 
\be
\ba{l}
\label{22sent}
e^{2i(\theta_3-\theta_2)}(1-is_3) (\sqrt{1-s_2^2 \eta_1} - is_2 \sqrt{1+\eta_1})(\sqrt{1-\nu^2 s_3^2 \eta_1} - i \nu s_3\sqrt{1+\eta_1}) = 
\\
\\
\hspace{2cm} =(1-is_2)(\sqrt{1-s_3^2 \eta_2} - is_3 \sqrt{1+\eta_2})(\sqrt{1-\nu^2 s_3^2 \eta_2} - i \nu s_3\sqrt{1+\eta_2})\;,
\ea
\ee
where the angular positions of the  vertices defined by equations \eqref{vertcoord11} are given by 
simple functions $\eta_{1,2} = \eta_{1,2}(s_2,s_3|\epsilon, \tilde \epsilon)$ along with 
$s_1 = s_2 +\nu \varkappa s_3$, where $\nu = \epsilon_3/\tilde \epsilon_1$ and $ \varkappa = \tilde \epsilon_1/\epsilon_1$ (see \cite{Alkalaev:2015wia} for more details).
In this form the second angular equation  depends on $s_2$ and $s_3$ only. Using the coordinate change \eqref{change2} the above equation can be rewritten as irrational equation $f(X,Y)=0$. Therefore, combining it with \eqref{321} we arrive at a couple of algebraic equations, quadratic and irrational ones which should have a common root with the monodromic equations \eqref{32}. By analogy with \eqref{512} we expect that of all  roots of equation \eqref{22sent} considered as $f(X,Y)=0$ we shall find a root $X = X(Y)$ which is to be treated as the second monodromic equation \eqref{32}.

Getting rid of the radicals in the second angular equation produces some higher order polynomial equation which defies exact solution. Using the super-light approximation we expand the  angular equations in small parameter $\epsilon_3$ around the seed solutions obtained by setting $\epsilon_3 = 0$ in  \eqref{321} and \eqref{22sent}. The expansion of angular momenta up to the third order  is given by 
\be
\label{1r}
\ba{l}
s_2=s_2^{(0)}+\nu  s_2^{(1)} + \nu^2 s_2^{(2)} +  \nu^3  s_2^{(3)} +...\;,
\\
\\
s_3=s_3^{(0)}+\nu s_3^{(1)} + \nu^2 s_3^{(2)} +...\;,
\ea
\ee
where  we used for convenience a rescaled deformation parameter $\nu = \epsilon_3/\tilde \epsilon_1$. Note that the above power series expansions have different highest orders in $\epsilon_3$. This is because the angular parameter $s_3$ enters the left-hand-side of the second equation in \eqref{Swvar} with $\epsilon_3$ prefactor so that to have the mechanical action $S = S(w_2,w_3)$ up to terms of order $\epsilon_3^{m}$ it is sufficient to have $s_2$ and $s_3$ up to terms of order $\epsilon_3^m$ and $\epsilon_3^{m-1}$, respectively. \footnote{In principle, this fact must follow directly from the geodesic length formula \eqref{actionFULL1}-\eqref{LLLLL}. We explicitly checked that substituting $s_3=s_3^{(0)}+\nu s_3^{(1)} + \nu^2 s_3^{(2)} + \nu^3 s_3^{(3)}+...$ into the length formula gives the same answer in the third order in $\nu$ exactly as if one sets $s_3^{(3)} = 0$.} 

The expansion coefficients are found to be  
\be
\ba{l}
\dps
s_2^{(0)}=- \cot\theta_2 + \varkappa \frac{1}{2 \sin\theta_2}\;,
\qquad 
s_2^{(1)}= \frac{\varkappa}{2}\cot(2\theta_3-\theta_2)\;,
\\
\\
\dps
s_2^{(2)}=\varkappa\frac{ [9 \cos(2 \theta_3)+ 7 \cos(2 \theta_2 - 2\theta_3) - \cos(4 \theta_2 - 6 \theta_3) + \cos(2 \theta_2 - 6 \theta_3) - 
   4 \cos(2 \theta_2 - 4 \theta_3) -12 ]}{32\sin^3( \theta_2 - 2 \theta_3)} \;,
\\
\\
\dps
s_2^{(3)}=\varkappa\frac{\sin\theta_3[\sin(\theta_2 - 3 \theta_3) - 3 \sin(\theta_2 - \theta_3)] [3 + \cos(2\theta_2 - 4 \theta_3) - 2 \cos(2\theta_2 - 2\theta_3) -  2 \cos(2 \theta_3)]}{8\sin^5(\theta_2 - 2 \theta_3)}\;,
\label{s2list}
\ea
\ee
and \footnote{In \cite{Alkalaev:2015wia} we used another but equivalent representation of the first expansion coefficient $s_3^{(1)}$. }
\be
\ba{l}
\dps
s_3^{(0)}=- \cot(2 \theta_3 - \theta_2)\;,
\qquad s_3^{(1)}=\frac{1}{2}\csc^3(\theta_2 - 2 \theta_3) [\sin^2\theta_2 + 4 \sin^2(\theta_2 - \theta_3) \sin^2\theta_3] \;,
\\
\\
\dps
s_3^{(2)}=-\frac{1}{16}\csc^5(\theta_2 - 2 \theta_3))[6 \cos\theta_2 + \cos(\theta_2 - 4 \theta_3) + \cos(3 \theta_2 - 4 \theta_3) - 
     8 \cos(\theta_2 - 2 \theta_3)] \times
\\
\\
\dps
\hspace{1cm} \times[3 + \cos(2\theta_2 - 4 \theta_3) - 2 \cos(2\theta_2 - 2\theta_3) -  2 \cos(2 \theta_3)] \;.
\label{s3list}
\ea
\ee
One can also  check that the consistency condition \eqref{conscon} rewritten in terms of variables $\theta_2$ and $\theta_3$ as   
$\epsilon_2 \partial s_2/\partial \theta_3=\epsilon_3 \partial s_3/\partial \theta_2$ is satisfied. 
To conclude, we note that in each order in $\nu$ or equivalently $\epsilon_3$ angular momenta corrections \eqref{s2list}-\eqref{s3list} are related to the accessory parameters corrections \eqref{xlist}-\eqref{ylist} according to the general relation \eqref{cs}. In this way we have shown  that the monodromic equation system does have the same root as the geodesic equation system.

\section{Classical conformal block vs  multi-particle action}
\label{sec:block}

Below we present the resulting expressions for the classical heavy-light conformal block obtained on the boundary and in the bulk  using respectively the monodromy and the geodesic approaches. The results obtained are in agreement  with the computation by means of the AGT combinatorial representation \cite{Alday:2009aq} performed for the lower level coefficients of the coordinate expansion of the classical conformal block. The corresponding  expressions are collected in the Appendix \bref{sec:appendix}.

\subsection{Classical conformal block}
The power series expansion of the 5-point classical conformal block  $f(z)$  \eqref{ccb} is given by 
\be
\label{exf}
f(z)=f^{(0)}(z)+\epsilon_3 f^{(1)}(z) + \epsilon_3^2 f^{(2)}(z)+ \epsilon_3^3 f^{(3)}(z) +...\;.
\ee
 Using explicit expressions for the accessory parameters \eqref{origch} and \eqref{xlist}-\eqref{ylist} 
 and integrating \eqref{cbgrad2} we find that expansion coefficients in \eqref{exf} are given by 
\be
\label{fcoef}
\ba{c}
\dps
f^{(0)}=-2\epsilon_1 \ln\big[\frac{a-1}{2 \sqrt{a}}\big]+\tilde{\epsilon}_1\ln\big[\frac{\sqrt{a}-1}{\sqrt{a}+1}\big]-\frac{\epsilon_1}{\alpha} \ln a\;,
\qquad 
f^{(1)}=-\ln\big[\frac{a - b^2}{2 \sqrt{a} b}\big]-\frac{1}{\alpha}\ln b\;,
\\
\\
\dps
f^{(2)}=-\frac{(a+b^2)(a+a^2-4 a b + b^2+ a b^2)}{4 \tilde{\epsilon}_1 \sqrt{a} (a-b^2)^2}\;,
\quad
f^{(3)}=\frac{( b-1) b (a - b) (a + b^2) (a + a^2 - 4 a b + b^2 + a b^2)}{2 \tilde{\epsilon}_1^2 (a-b^2)^4}\;,
\ea
\ee
where $a = (1-z_2)^\alpha$ and $b = (1-z_3)^\alpha$, see  \eqref{origch}. The leading contribution $f^{(0)}$  is the $4$-point classical heavy-light conformal block \cite{Fitzpatrick:2014vua,Hijano:2015rla}. 

\subsection{Multi-particle action}

The power series expansion of the bulk multi-particle action $S(w)$  \eqref{actionFULL1} is given by 
\be
\label{exS}
S(w)=S^{(0)}(w)+\epsilon_3 S^{(1)}(w) + \epsilon_3^2 S^{(2)}(w)+ \epsilon_3^3 S^{(3)}(w) +...\;.
\ee
Using explicit expressions for the angular momenta \eqref{s2list} - \eqref{s3list} and integrating \eqref{Swvar} we find that expansion coefficients in \eqref{exS} are given by 
\be
\label{Scoef}
\ba{c}
\dps
S_0(\theta)=  - 2 \epsilon_1 \ln\sin \theta_2 + \tilde\epsilon_1  \ln\tan\frac{\theta_2}{2}\;,
\qquad
S_1(\theta)=-  \ln \sin(2 \theta_3-\theta_2) \;,
\\
\\
\dps
S_2(\theta)=-\frac{\cos\theta_2 + 2 \csc^2(\theta_2 - 2 \theta_3) \sin(\theta_2 - \theta_3) \sin\theta_3}{2 \tilde{\epsilon}_1}\;,
\\
\\
\dps
S_3(\theta)=-\frac{\cos\theta_2 + 2 \csc^2(\theta_2 - 2 \theta_3) \sin(\theta_2 - \theta_3) \sin\theta_3}{2 \tilde{\epsilon}_1}\times
\frac{4 \csc^2(\theta_2 - 2 \theta_3) \sin(\theta_2 - \theta_3) \sin\theta_3}{2 \tilde{\epsilon}_1}\;,
\ea
\ee
where we switched to $\theta_2$ and $\theta_3$, see \eqref{foot1}. Expansion coefficients \eqref{Scoef} are related to  \eqref{fcoef} according to the general identification formula  \eqref{fS}. The same results  follow from the explicit geodesic length formula \eqref{actionFULL1}-\eqref{LLLLL}. We note that \eqref{Swvar} as well as \eqref{cbgrad2} allow to define the action (the conformal block) up to some coordinate independent constant, while explicit expressions \eqref{actionFULL1}-\eqref{LLLLL} allow to fix the constant.

\section{Conclusion}
\label{sec:concl}

We have computed the 5-point heavy-light conformal block in the  super-light approximation up to the third order with respect to  the conformal 
dimension of one of the three light fields. The computation has been done in two independent ways: using the monodromy and  the geodesic approaches. 
The resulting expressions coincide. 
We observe different aspects of the correspondence between the two methods.
In particular, we find that the boundary variables and equations find their counterparts in the bulk consideration. There is also a precise relation between the accessory parameters and the conserved angular momenta of the different geodesic segments. 

The similarity between bulk and boundary computations leads to the natural
assumption that in the present  context the AdS$_3$/CFT$_2$ correspondence is to be understood  in a strong sense, \textit{i.e.} as two different descriptions of the same Liouville theory.
Indeed, in the semiclassical limit the correlators of the primary fields are dominated  by the stationary-point configuration which is given by the Liouville equation with the delta-function sources corresponding to the heavy fields. The light fields do not affect the stationary-point  configuration and are considered as propagating in the background geometry 
formed by the heavy ones. In our case there are two heavy operators with equal conformal weights so that we are dealing with the dynamics of the light fields propagating in the geometry of the  Poincar\'{e}  disk with two singularities inside.  
One can see that this geometry is
nothing other than the two-dimensional slice geometry of the angle deficit/BTZ black hole in three dimensions. In particular, we have seen that the geodesic method deals with a constant time or angle slices which are Poincar\'{e} disks.   

This link can be traced also in the opposite direction.
Indeed, provided the Brown-Henneaux boundary condition is satisfied, the three-dimensional gravity with the negative cosmological constants is described by the Liouville theory \cite{Coussaert:1995zp}. Another way to see this theory inside the three-dimensional gravity  is   to represent the bulk metric as a stack of the Poincar\'{e} disks so that the exponential scale factor  satisfies the residual Einstein equation which is again the Liouville equation \cite{Welling:1997fw}. \footnote{The above Einstein/Liouville relation  conforms with the well-known correspondence between two-dimensional Virasoro  conformal  blocks and physical states in three-dimensional quantum $SL(2, \mathbb{C})$ Chern-Simons theory  \cite{Verlinde:1989ua}. See, \textit{e.g.}, refs. \cite{Carlip:2005zn,Krasnov:2000zq,Jackson:2014nla,Verlinde:2015qfa,Kim:2015qoa} for more discussion of  quantum   $3d$ gravity and the Liouville theory.} Adding light particles moving in the background geometry generated by static heavy sources one arrives at the Liouville equation with the heavy source terms supplemented with the standard geodesic equations for interacting  light particles. This corresponds to decomposing the semiclassical correlation functions into conformal blocks and introducing intermediate channels.

Following the above heuristic arguments we tend to conclude that the Lagrangian description of the classical Liouville theory has to allow for establishing literal matching of the boundary and the bulk descriptions in the semiclassical limit. Current developments strongly indicate that this is indeed the case but an exact method  of deriving the bulk geodesic description starting from the semiclassical Liouville theory is not yet available.

\vspace{4mm}

\noindent \textbf{Acknowledgements.} The work of K.A. was supported by RFBR grant No 14-01-00489. 
V.B. thanks Giuseppe Mussardo for hospitality at SISSA (Trieste) and support within the European Grant
Project QICFT.

\appendix

\section{The classical  block from the AGT representation}
\label{sec:appendix}

Here we compare the results of Section \bref{sec:block} with those coming from the semiclassical computation for  the combinatorial AGT representation \cite{Alday:2009aq} of the conformal blocks sketched  in \cite{Alkalaev:2015wia}.  Introducing  parameters $q_1$ and $q_2$ which define coordinates of the 5-point block as 
$z_2 = q_1 q_2$ and  $z_3 =q_2$ the classical conformal block in the $q$-representation is given by 
\begin{equation}
\label{5classblock}
\mathcal{\tilde F}(q_1,q_2)= e^{-\frac{c}{6} \tilde{f}(q_1,q_2)}\;,\qquad c\rightarrow \infty\;,
\end{equation}
where $c$ is the central charge and $\tilde \cF(q_1, q_2)$ is the conformal block in the AGT representation normalized as  $\tilde \cF(0)=1$ so that $\tilde f(0)=0$. We note that the classical conformal block in \eqref{ccb} has different asymptotic behavior: $f(z) \sim \log z$ at $z\rightarrow 0$.

Using the power series expansion of the classical conformal block in the new coordinates
\be
\label{AGTcoef}
\tilde f(q)=\tilde f^{(0)}(q)+\epsilon_3 \tilde f^{(1)}(q)+\epsilon_3^2 \tilde f^{(2)}(q)+\epsilon_3^3 \tilde f^{(3)}(q)+...\;,
\ee
and applying the AGT combinatorial computational scheme  we can find the expansion coefficients $\tilde f^{(k)}(q)$ in the from of   $q_1$ and $q_2$ power series  expansions. 
It is convenient to introduce an auxiliary parameter $t$ such that $t^m$ term takes into account all contributions
of the terms $q_1^{m_1}q_2^{m_2}$ with $m=m_1+m_2$. Up to $m=5$ we get 
\be
\label{f0}
\ba{l}
\dps
\tilde f^{(0)}(q_1,q_2|t)=-\frac{1}{2} (\tilde{\epsilon}_1 q_1 q_2) t^2 +\frac{1}{48}(
-4 \epsilon_1 q_1^2 q_2^2 - 10 \tilde{\epsilon}_1 q_1^2 q_2^2 + 4 \epsilon_1 q_1^2 q_2^2 \alpha^2 + 
 \tilde{\epsilon}_1  q_1^2 q_2^2 \alpha^2) t^4+
 \\
 \\
\dps
\hspace{2.5cm}+\frac{1}{48} (-4 \epsilon_1 q_1^3 q_2^3 - 6 \tilde{\epsilon}_1 q_1^3 q_2^3 + 4 \epsilon_1 q_1^3 q_2^3 \alpha^2 + 
   \tilde{\epsilon}_1 q_1^3 q_2^3 \alpha^2)t^6+
 \cO(t^7)\;,
\ea
\ee
\be
\label{f1}
\ba{l}
\dps
\tilde f^{(1)}(q_1,q_2|t)=-\frac{1}{2} (q_1+ q_2) t +\frac{1}{24}(-3 q_1^2 + 6 q_1 q_2 - 7 q_2^2 + 4 q_2^2 \alpha^2) t^2
+\frac{1}{24}(-q_1^3 - 3 q_1^2 q_2 +
 \\
 \\
\dps
\hspace{2.5cm}+ q_1 q_2^2 - 5 q_2^3 - 4 q_1 q_2^2 \alpha^2 + 
  4 q_2^3 \alpha^2)t^3+\frac{1}{2880} (-45 q_1^4 - 180 q_1^3 q_2 + 330 q_1^2 q_2^2+
 \\
 \\
\dps
\hspace{2.5cm} + 60 q_1 q_2^3 - 469 q_2^4 + 
  120 q_1^2 q_2^2 \alpha^2 - 240 q_1 q_2^3 \alpha^2 + 
  440 q_2^4 \alpha^2 - 16 q_2^4 \alpha^4) t^4+
 \cO(t^5)\;,
\ea
\ee
\be
\label{f2}
\ba{l}
\dps
\tilde f^{(2)}(q_1,q_2|t)=-\frac{1}{16 \tilde{\epsilon}_1} (q_1^2+ q_2^2\alpha^2) t^2 +
\frac{1}{16 \tilde{\epsilon}_1}(-q_1^3 + q_1^2 q_2 + q_1 q_2^2 \alpha^2 - q_2^3 \alpha^2) t^3
+ \frac{1}{192 \tilde{\epsilon}_1}(-9 q_1^4 +
 \\
 \\
\dps
\hspace{2.5cm} + 6 q_1^3 q_2 - q_1^2 q_2^2 - 11 q_1^2 q_2^2 \alpha^2 + 
  6 q_1 q_2^3 \alpha^2 - 11 q_2^4 \alpha^2 + 2 q_2^4 \alpha^4) t^4+\cO(t^5)\;,
\ea
\ee
\be
\label{f3}
\ba{l}
\dps
\tilde f^{(3)}(q_1,q_2|t)=\frac{1}{128 \tilde{\epsilon}_1^2} (-q_1^4 + 2 q_1^2 q_2^2 \alpha^2 - q_2^4 \alpha^4) t^4 +
\frac{1}{64 \tilde{\epsilon}_1^2}(-q_1^5 + q_1^4 q_2 + q_1 q_2^4 \alpha^4 - q_2^5 \alpha^4) t^5
+\cO(t^6)\;.
\ea
\ee
Setting $t=1$ yields expansion coefficients in the decomposition \eqref{AGTcoef}. 
We find out  that  up to coordinate  independent terms 
\be
S(\theta_2,\theta_3) = \tilde f(q_2,q_3) - \epsilon_1 \ln(1-q_1q_2) - \epsilon_3 \ln(1-q_2)
+ (2\epsilon_1  - \tilde \epsilon_1) \ln q_1 q_2 - \epsilon_3 \ln q_2  \;.
\ee
Recalling the difference in the asymptotic behavior of $\tilde f(q)$ and $f(z)$ (modulo the standard prefactor $z_2^{2\epsilon_1 - \tilde\epsilon_1} z_3^{-\epsilon_3} $) we see that the above identification conforms \eqref{fS}.

%

\begin{thebibliography}{10}

\bibitem{Belavin:1984vu}
A.~Belavin, A.~M. Polyakov, and A.~Zamolodchikov, ``{Infinite Conformal
  Symmetry in Two-Dimensional Quantum Field Theory},'' {\em Nucl.Phys.} {\bf
  B241} (1984)
333--380.

\bibitem{Fitzpatrick:2014vua}
A.~L. Fitzpatrick, J.~Kaplan, and M.~T. Walters, ``{Universality of
  Long-Distance AdS Physics from the CFT Bootstrap},'' {\em JHEP} {\bf 1408}
  (2014) 145,
\href{http://www.arXiv.org/abs/1403.6829}{{\tt 1403.6829}}.

\bibitem{Asplund:2014coa}
C.~T. Asplund, A.~Bernamonti, F.~Galli, and T.~Hartman, ``{Holographic
  Entanglement Entropy from 2d CFT: Heavy States and Local Quenches},'' {\em
  JHEP} {\bf 1502} (2015) 171,
\href{http://www.arXiv.org/abs/1410.1392}{{\tt 1410.1392}}.

\bibitem{Caputa:2014eta}
  P.~Caputa, J.~Simon, A.~Stikonas and T.~Takayanagi,
  JHEP {\bf 1501} (2015) 102
  doi:10.1007/JHEP01(2015)102
  [arXiv:1410.2287 [hep-th]].

\bibitem{Fitzpatrick:2015zha}
A.~L. Fitzpatrick, J.~Kaplan, and M.~T. Walters, ``{Virasoro Conformal Blocks
  and Thermality from Classical Background Fields},''
\href{http://www.arXiv.org/abs/1501.05315}{{\tt 1501.05315}}.

\bibitem{Hijano:2015rla}
E.~Hijano, P.~Kraus, and R.~Snively, ``{Worldline approach to semi-classical
  conformal blocks},'' {\em JHEP} {\bf 07} (2015) 131,
\href{http://www.arXiv.org/abs/1501.02260}{{\tt 1501.02260}}.

\bibitem{Alkalaev:2015wia}
K.~B. Alkalaev and V.~A. Belavin, ``{Classical conformal blocks via AdS/CFT
  correspondence},'' {\em JHEP} {\bf 08} (2015) 049,
\href{http://www.arXiv.org/abs/1504.05943}{{\tt 1504.05943}}.

\bibitem{Hijano:2015qja}
E.~Hijano, P.~Kraus, E.~Perlmutter, and R.~Snively, ``{Semiclassical Virasoro
  Blocks from AdS$_3$ Gravity},''
\href{http://www.arXiv.org/abs/1508.04987}{{\tt 1508.04987}}.

\bibitem{Zamolodchikov:1995aa}
A.~B. Zamolodchikov and A.~B. Zamolodchikov, ``{Structure constants and
  conformal bootstrap in Liouville field theory},'' {\em Nucl.Phys.} {\bf B477}
  (1996) 577--605,
\href{http://www.arXiv.org/abs/hep-th/9506136}{{\tt hep-th/9506136}}.

\bibitem{Hijano:2015zsa}
E.~Hijano, P.~Kraus, E.~Perlmutter, and R.~Snively, ``{Witten Diagrams
  Revisited: The AdS Geometry of Conformal Blocks},''
\href{http://www.arXiv.org/abs/1508.00501}{{\tt 1508.00501}}.

\bibitem{Zamolodchikov:1985ie}
A.~Zamolodchikov, ``{Conformal symmetry in two-dimensions: an explicit
  recurrence formula for the conformal partial wave amplitude},'' {\em
  Commun.Math.Phys.} {\bf 96} (1984)
419--422.

\bibitem{Zamolodchikov:1987ie}
A.~Zamolodchikov, ``{Conformal Symmetry in Two-dimensional Space: Recursion
  Representation of the Conformal Block},'' {\em Teor.Mat.Fiz.} {\bf 73} (1987)
  103--110.

\bibitem{Perlmutter:2015iya}
E.~Perlmutter, ``{Virasoro conformal blocks in closed form},'' {\em JHEP} {\bf
  08} (2015) 088,
\href{http://www.arXiv.org/abs/1502.07742}{{\tt 1502.07742}}.

\bibitem{Fitzpatrick:2015foa}
A.~L. Fitzpatrick, J.~Kaplan, M.~T. Walters, and J.~Wang, ``{Hawking from
  Catalan},''
\href{http://www.arXiv.org/abs/1510.00014}{{\tt 1510.00014}}.

\bibitem{Alday:2009aq}
L.~F. Alday, D.~Gaiotto, and Y.~Tachikawa, ``{Liouville Correlation Functions
  from Four-dimensional Gauge Theories},'' {\em Lett.Math.Phys.} {\bf 91}
  (2010) 167--197,
\href{http://www.arXiv.org/abs/0906.3219}{{\tt 0906.3219}}.

\bibitem{Bershtein:2014qma}
M.~Bershtein and O.~Foda, ``{AGT, Burge pairs and minimal models},'' {\em JHEP}
  {\bf 1406} (2014) 177,
\href{http://www.arXiv.org/abs/1404.7075}{{\tt 1404.7075}}.

\bibitem{Alkalaev:2014sma}
K.~Alkalaev and V.~Belavin, ``{Conformal blocks of $W_N$ minimal models and AGT
  correspondence},'' {\em JHEP} {\bf 1407} (2014) 024,
\href{http://www.arXiv.org/abs/1404.7094}{{\tt 1404.7094}}.


\bibitem{Balasubramanian:2014sra}
  V.~Balasubramanian, B.~D.~Chowdhury, B.~Czech and J.~de Boer,
  JHEP {\bf 1501} (2015) 048, 
  \href{http://www.arXiv.org/abs/1406.5859}{{\tt 1406.5859}}.


\bibitem{Zamolodchikov1986}
A.~Zamolodchikov, ``{Two-dimensional conformal symmetry and critical four-spin
  correlation functions in the Ashkin-Teller model},'' {\em Zh. Eksp. Teor.
  Fiz.} {\bf 90} (1986) 1808--1818.

\bibitem{Harlow:2011ny}
D.~Harlow, J.~Maltz, and E.~Witten, ``{Analytic Continuation of Liouville
  Theory},'' {\em JHEP} {\bf 1112} (2011) 071,
\href{http://www.arXiv.org/abs/1108.4417}{{\tt 1108.4417}}.

\bibitem{Zograf}
L.~A. Takhtajan and P.~Zograf, {\em {Funct. An. Appl.}} {\bf 19} (1986) 219.

\bibitem{Takhtajan:1994vt}
L.~A. Takhtajan, ``{Topics in quantum geometry of Riemann surfaces:
  Two-dimensional quantum gravity},'' in {\em {Como Quantum Groups
  1994:541-580}}, pp.~541--580.
\newblock 1994.
\newblock
\href{http://www.arXiv.org/abs/hep-th/9409088}{{\tt hep-th/9409088}}.
\newblock

\bibitem{Coussaert:1995zp}
O.~Coussaert, M.~Henneaux, and P.~van Driel, ``{The Asymptotic dynamics of
  three-dimensional Einstein gravity with a negative cosmological constant},''
  {\em Class. Quant. Grav.} {\bf 12} (1995) 2961--2966,
\href{http://www.arXiv.org/abs/gr-qc/9506019}{{\tt gr-qc/9506019}}.

\bibitem{Welling:1997fw}
M.~Welling, ``{Explicit solutions for point particles and black holes in spaces
  of constant curvature in (2+1)-Dimensional gravity},'' {\em Nucl.Phys.} {\bf
  B515} (1998) 436--452,
\href{http://www.arXiv.org/abs/hep-th/9706021}{{\tt hep-th/9706021}}.

\bibitem{Verlinde:1989ua}
H.~L. Verlinde, ``{Conformal Field Theory, 2-$D$ Quantum Gravity and
  Quantization of Teichmuller Space},'' {\em Nucl. Phys.} {\bf B337} (1990)
652.

\bibitem{Carlip:2005zn}
S.~Carlip, ``{Conformal field theory, (2+1)-dimensional gravity, and the BTZ
  black hole},'' {\em Class. Quant. Grav.} {\bf 22} (2005) R85--R124,
\href{http://www.arXiv.org/abs/gr-qc/0503022}{{\tt gr-qc/0503022}}.

\bibitem{Krasnov:2000zq}
K.~Krasnov, ``{Holography and Riemann surfaces},'' {\em Adv. Theor. Math.
  Phys.} {\bf 4} (2000) 929--979,
\href{http://www.arXiv.org/abs/hep-th/0005106}{{\tt hep-th/0005106}}.

\bibitem{Jackson:2014nla}
S.~Jackson, L.~McGough, and H.~Verlinde, ``{Conformal Bootstrap, Universality
  and Gravitational Scattering},''
\href{http://www.arXiv.org/abs/1412.5205}{{\tt 1412.5205}}.

\bibitem{Verlinde:2015qfa}
H.~Verlinde, ``{Poking Holes in AdS/CFT: Bulk Fields from Boundary States},''
\href{http://www.arXiv.org/abs/1505.05069}{{\tt 1505.05069}}.

\bibitem{Kim:2015qoa}
J.~Kim and M.~Porrati, ``{On a Canonical Quantization of 3D Anti de Sitter Pure
  Gravity},''
\href{http://www.arXiv.org/abs/1508.03638}{{\tt 1508.03638}}.

\end{thebibliography}
\providecommand{\href}[2]{#2}\begingroup\raggedright\endgroup

\end{document}